\documentclass[prl,twocolumn,fleqn]{revtex4}
\usepackage{amscd,verbatim}
\usepackage[all]{xy}
\usepackage{graphicx}

\usepackage{amsmath}
\usepackage[psamsfonts]{amssymb}
\usepackage{mathtools}
\usepackage{euscript}

\usepackage{latexsym}

\renewcommand{\(}{\begin{equation}}
\renewcommand{\)}{end{equation} \vspace{-.05in}\linebreak}

\newcounter{saveeqn}
\newcounter{savealpheqn}

\newcommand{\alpheqn}{\setcounter{saveeqn}{\value{equation}}%
  \stepcounter{saveeqn}\setcounter{equation}{0}%
  \renewcommand{\theequation}{\mbox{\arabic{section}.\arabic{saveeqn}
\alph{equation}}}
  \renewcommand{\)}{\end{equation}}}
\def\part#1{\frac{\partial}{\partial{#1}}}%
\def\group#1{\refstepcounter{equation}\setcounter{saveeqn}
 {\value{equation}}%
  \label{#1}\setcounter{equation}{0}%
\renewcommand{\theequation}{\mbox{\arabic{section}.\arabic{saveeqn}
\alph{equation}}}
  \renewcommand{\)}{\end{equation}}}
\newcommand{\reseteqn}{\setcounter{equation}{\value{saveeqn}}%
  \renewcommand{\theequation}{\arabic{section}.\arabic{equation}}%
  \renewcommand{\)}{\end{equation}}}

\newcommand{\aalpheqn}{\setcounter{saveeqn}{\value{equation}}%
  \stepcounter{saveeqn}\setcounter{equation}{0}%
  \renewcommand{\theequation}{\mbox{
        \Alph{subsection}.\arabic{saveeqn}\alph{equation}}}
   \renewcommand{\)}{\end{equation}}}
\newcommand{\areseteqn}{\setcounter{equation}{\value{saveeqn}}%
  \renewcommand{\theequation}{\Alph{subsection}.\arabic{equation}}%
  \renewcommand{\)}{\end{equation}}}

\renewcommand{\thefootnote}{\alph{footnote}}
\renewcommand{\(}{\begin{equation}}
\renewcommand{\)}{\end{equation}}
\newcommand{\ba}{\begin{eqnarray}}
\newcommand{\ea}{\end{eqnarray}}

\newcommand{\bp}{\mathop{\vtop{\ialign{##\crcr
   $\hfil\displaystyle{}\hfil$\crcr\noalign{\kern-13pt\nointerlineskip}
   \BIG{(}\hskip0pt\crcr\noalign{\kern3pt}}}}}
\newcommand{\cbp}{\mathop{\vtop{\ialign{##\crcr
   $\hfil\displaystyle{}\hfil$\crcr\noalign{\kern-13pt\nointerlineskip}
   \BIG{)}\hskip0pt\crcr\noalign{\kern3pt}}}}}
\newcommand{\pa}{\mathop{\vtop{\ialign{##\crcr

$\hfil\displaystyle{\oplus}\hfil$\crcr\noalign{\kern+1pt\nointerlineskip
}
   \hspace{.08in}$^{\alpha=0}$\hskip6pt\crcr\noalign{\kern3pt}}}}}

\newcommand{\Z}{\ensuremath{\mathbb Z}}

\newcommand{\beq}{\begin{equation}}
\newcommand{\eeq}{\end{equation}}

\numberwithin{equation}{section}
\renewcommand{\theequation}{\mbox{\arabic{equation}}}

\catcode`\@=11
\def\vereq#1#2{\lower3pt\vbox{\baselineskip1.5pt \lineskip1.5pt
\ialign{$\m@th#1\hfill##\hfil$\crcr#2\crcr\sim\crcr}}}
\catcode`\@=12

\makeatletter
\newcommand\figcaption{\def\@captype{figure}\caption}
\newcommand\tabcaption{\def\@captype{table}\caption}
\makeatother
\renewcommand{\(}{\begin{equation}}
\renewcommand{\)}{\end{equation}}

\newcommand{\Exterior}{\scalebox{.8}{\ensuremath \bigwedge}}

\begin{document}
\def\thefootnote{\fnsymbol{footnote}}

\title{Higher T-duality in M-theory via local supersymmetry}

\author{Hisham Sati$^1$ and Urs Schreiber$^{1,2}$}
\affiliation{
$^1$ Division of Science and Mathematics, New York University, Abu Dhabi, UAE.\\
$^2$ Czech Academy of Sciences, Mathematics Institute, {\v Z}itna 25, 115 67 Praha 1, Czech Republic.}

\begin{abstract}
By analyzing super-torsion and brane super-cocycles,
we derive a new duality in M-theory, which takes the form of a higher
version of T-duality in string theory. This involves
a new topology change mechanism abelianizing the 3-sphere associated with the C-field topology
to the 517-torus associated with exceptional-generalized super-geometry.
Finally we explain parity symmetry in M-theory within exceptional-generalized super-spacetime at the same level of
spherical T-duality, namely as an isomorphism on 7-twisted cohomology.
\end{abstract}

%
\setcounter{footnote}{0}
\renewcommand{\thefootnote}{\arabic{footnote}}


\maketitle

\renewcommand{\thepage}{\arabic{page}}

One of the fundamental open question in QFT and in string theory is formulating
the non-perturbative theory.
Since string theory seems to be the more constrained of the two (see \cite{Vafa05}), its non-perturbative
effects are more constrained, related to \emph{$p$-branes} in the theory \cite{Polchinski94, BeckerBeckerStrominger95},
governed by a web of subtle equivalences, called dualities,
notably \emph{T-dualities} (\cite{Bu1, Bu2}, see \cite{GPR}).
These are indicative of an underlying non-perturvative structure, namely
 \emph{M-theory} (see \cite{Duff99}).
Certain well-known dualities do have an M-theoretic origin, e.g. \cite{GV1, GV2}.

Here, replacing the string with the M5-brane, we uncover a duality fully within M-theory,
which may be formulated as a higher-structural analog of T-duality in string theory.
We derive this using constraints on super $p$-brane fluxes due to local supersymmetry (supergravity)
\cite{AETW87, AzTo89}
and use recent rigorous results of \cite{Higher}, which generalize a
recent derivation \cite{FSS16} of the brane flux sector T-duality,
previously proposed as ``topological T-duality'' \cite{BEM}.

Our derivation is rooted in the fundamental
super-torsion constraint of 11d supergravity \cite{CL, Howe97}.
This says that, while the world need not exhibit global large scale supersymmetry,
each tiny (mathematically: infinitesimal) neighborhood of any event in spacetime is
supersymmetric.
Moreover, and this is key to our analysis, the fermionic component of the brane charges (fluxes)
restricted to any of these infinitesimal neighborhoods is constrained to be a non-trivial solution
to the supersymmetric Gauss law (mathematically: a non-trivial cocycle in the Chevalley-Eilenberg
CE-complex of the supersymmetry algebra) -- see \cite{BST1, BST2, BLNPST97}.

\vspace{-3mm}
\begin{center}
\hspace{-5mm}
 \includegraphics[width=.51\textwidth]{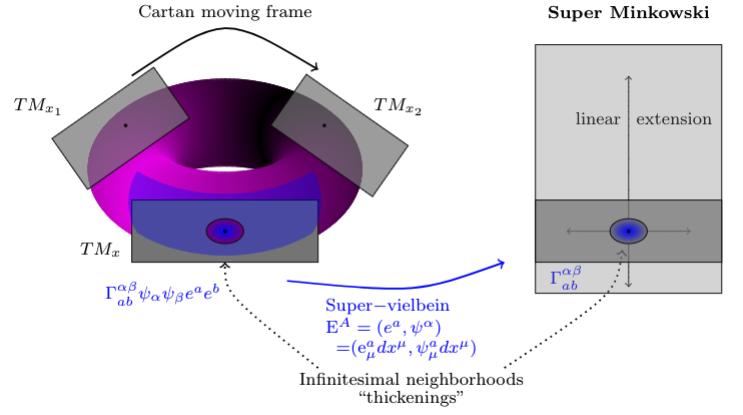}
\end{center}

\vspace{-2mm}
\noindent The full implication of these constraints has perhaps not been fully appreciated
until recently.
Indeed, in \cite{FSS16} we showed that these constraints already imply the Buscher
rules \cite{Bu1, Bu2} for the F1/D$p$-brane charge sector of
T-duality.

Turning this around, it means that analysis of these local supersymmetry constraints may be
 used to systematically discover and analyze previously unknown facts about
M-theory.
This way, we find in equations \eqref{eC6Twisted} and \eqref{IsoIn7Twisted} below the
 duality of M5-branes exchanging 3-spherical wrapping modes with
non-wrapping modes, in higher analogy to how ordinary T-duality exchanges winding modes
of strings within D-branes.

\vspace{-2mm}
\begin{center}
 \includegraphics[width=.12\textwidth]{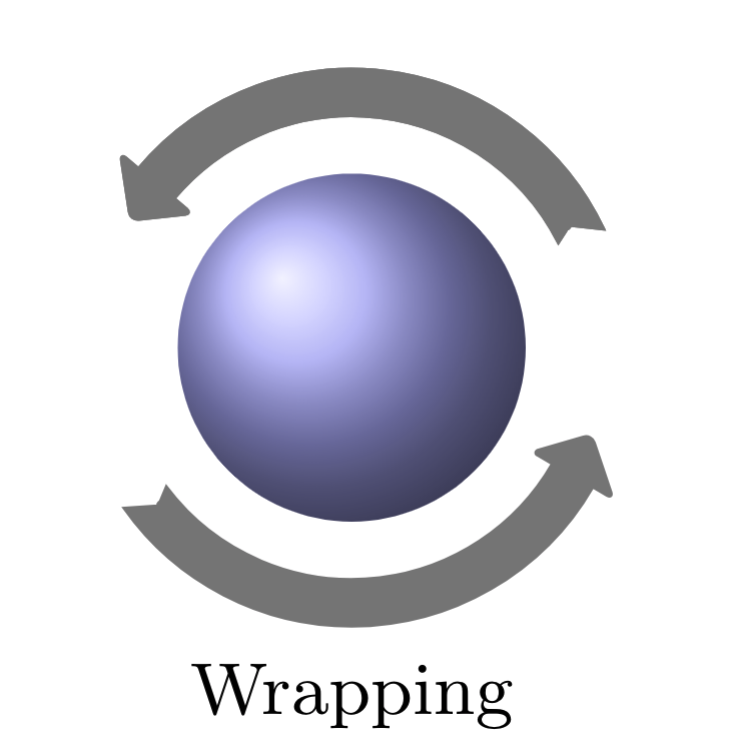} \qquad
 \includegraphics[width=.12\textwidth]{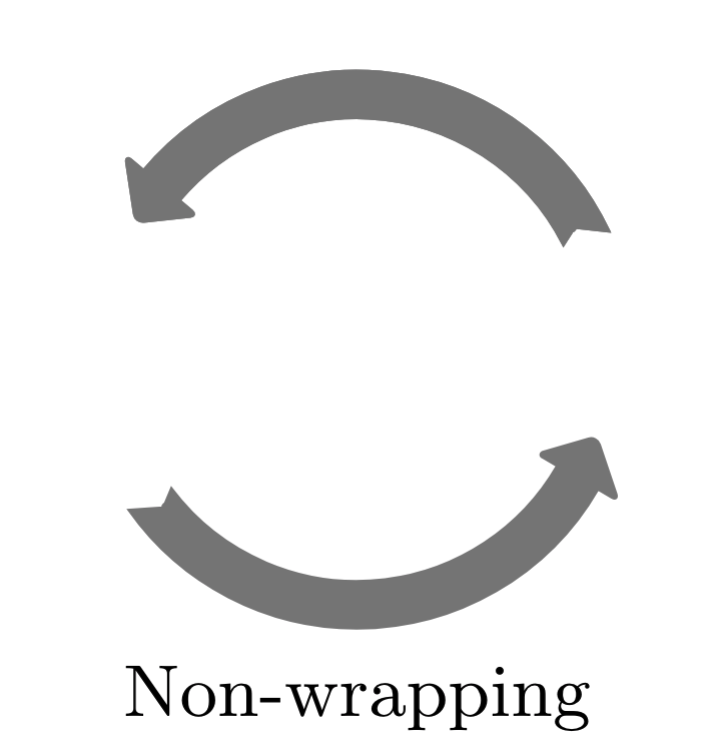}
\end{center}
This yields an equivalence in 7-twisted cohomology, as a special case of the general theory of \cite{LSW}, the twist arising from the M5-brane flux.
The physical picture is similar to -- but subtly different from --  a
recent proposal for ``spherical topological T-duality" \cite{BEM2}, in that the 3-sphere
is not part of spacetime itself, but of extended spacetime, as established in \cite{FSS13}.
The higher T-duality of M5-branes is related to the fact that the flux for the M5-brane
\emph{necessarily} has an algebraic form that is a higher-degree analog of that of a T-dualizable
H-flux in string theory; see equation \eqref{C3-T} below.

In fact, we observe that the 3-spherical fiber over spacetime may be traded in for a 517-dimensional torus bundle
which turns out to realize the \emph{exceptional-generalized} geometry in M-theory previously proposed by Hull
\cite{Hull07b}.
\begin{center}
 \includegraphics[width=.5\textwidth]{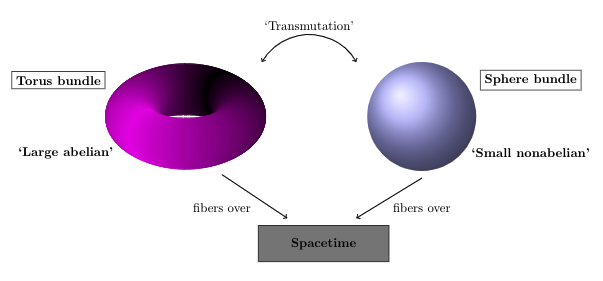}
\end{center}
The subtle nature of these 3-spherical wrapping modes can be explained in terms of
a new ``C-cohomology'' of the C-field on exceptional-generalized spacetime, see \eqref{CCohomology}.

We emphasize that the new duality which we present is of same nature as the ``topological T-duality''
of \cite{BEM}, just in higher degree. In fact, it is of the form of ``spherical topological T-duality'' in the sense of \cite{BEM2, LSW},
except that we clarify that the spherical fiber is either that of the C-field 2-gerbe \emph{over} spacetime, or a subtle kind of image of that in
the exceptional generalized tangent bundle. This means that it is a duality (just) of the brane charges (whence ``topological duality''), incarnated as an equivalence
of the corresponding twisted cohomology groups (see \eqref{IsoIn7Twisted} below). We highlight that the ``topological T-duality'' of \cite{BEM} was motivated from string theory,
but not explicitly derived from it (see \cite{Hull07a} for discussion), until Ref. \cite{FSS16} showed that its postulated axioms
indeed follow from analysis of the superspace torsion constraints of the F1/D$p$ fluxes. Here, along with \cite{Higher}, we perform this kind of superspace analysis, but now for the M-brane fluxes instead of D-brane fluxes. We expect that a local geometric version of
the spherical T-duality in M-theory, in which the metric appears explicitly, may be derived analogously, using supersymmetric Cartan moving frames, as indicated above.

In this letter, we report on the following:

\noindent {\bf 1.} A new duality in M-theory:  3-spherical T-duality for M5-branes.

\noindent {\bf 2.}   A new topology change mechanism: transmutation of 3-spherical fiber to a 517-torus fiber.

\noindent {\bf 3.}   Insight on understanding of the role of exceptional generalized super-geometry of this torus.

\noindent {\bf 4. } Lifting of parity symmetry on exceptional generalized spacetime to the level of
spherical T-duality.

{\it We also emphasize that this is a derivation of M-theoretic structures from first principles, not involving
any extrapolation from perturbative string theory nor any conjectures or informal analogies from other sources.}

\section{\bf  A new duality in M-theory}

It turns out that the phenomenon of ordinary T-duality, when restricted to its action on
brane charges (topological T-duality) is, at its heart, encoded in the peculiar algebraic form of a
T-dualizable H-flux:
\(
\label{H3-T}
  \tilde H_3 \;=\; H_3 + \theta \wedge F'_2
  \,.
\)
Here $H_3$ is the  basic H-flux on the base space over which T-duality takes place, $\theta$ is the Maurer-Cartan form
on the torus bundle over that base space which is being T-dualized, and $F'_2$ is the first Chern form of the
\emph{other}, namely the T-dual torus bundle (an index contraction over torus directions is left implicit).
Relation \eqref{H3-T} equivalently means that integrating the H-flux over T-dualized directions yields the T-dual Chern class:
\(
  \label{H3-Push}
  \int_T  \tilde H = F'_2
  \,,
\)
which is the condition proposed in \cite{BEM}.

Now the 11-dimensional Chern-Simons term in 11d supergravity implies that the 7-form flux
$\tilde{G}_7$ to which
 the  M5-brane couples is of a similar algebraic form, but in higher degree, controlled by the corresponding
supercocylce $\tilde \mu_{M5}$ \cite{FSS15}:
\(
\label{C3-T}
  \begin{array}{ccccc}
 \tilde \mu_{M5}
    &=&
    \tfrac{1}{5!} \overline{\psi} \Gamma_{I_5} \psi \wedge e^{I_5}
    &+&
    c_3 \wedge \tfrac{i}{2}
    \overline{\psi} \Gamma_{I_2} \psi \wedge e^{I_2}
    \\
 \tilde G_7
    &=&
    G_7 &+& \tfrac{1}{2} C_3 \wedge G_4
  \end{array}
\)
Here $\Gamma_{I_m}$ and $e^{I_m}$ are shorthand for
$\Gamma_{a_1 \cdots a_m}$ and $e^{a_1} \wedge \cdots \wedge e^{a_m}$,
respectively.
In the higher generalization of \eqref{H3-Push}  this means that upon integration we have
\(
  \int_{S^3} \tilde G_7  = G_4\;.
\)
A subtlety here is that, contrary to the Maurer-Cartan form $\theta$ in \eqref{H3-T}, the C-field $C_3$
in \eqref{C3-T} is not in general globally defined on 11-dimensional spacetime. It does, however,
become globally defined on a higher-dimensional bundle \emph{over} spacetime. Strictly speaking,
this is a higher bundle structure called a \emph{2-gerbe} or \emph{$B^2 U(1)$-principal bundle}.
There, as a first approximation, one may retain only those global topological effects that may be
detected by differential form data alone, hence by ``higher Gauss laws'' (mathematically this means
to work in ``rational homotopy theory''). Then this C-field 2-gerbe is equivalently just a 3-sphere
bundle over 11-dimensional spacetime. In a dual incarnation, we can re-interpret this rational 3-sphere bundle
to have played a crucial role in the elegant re-derivation of 11-dimensional supergravity in
\cite{DF}.

This reveals that the C-field $C_3$ may be thought of as a higher degree analog of the
Maurer-Cartan 1-form $\theta$ of a torus bundle, as indicated in the  table below.
This is sufficient to deduce that the logic of topological T-duality applies,  revealing
that there is a higher analog on the 3-sphere bundle over 11-dimensional super-spacetime
of what in string theory is Hori's formula \cite{Hori} for the Buscher rules on RR-fields. The latter
sends a D-brane flux $e^{B_2} \wedge \mathcal{C}$ to its
T-dual flux given by  (see \cite{T})
\(
  e^{B'_2} \wedge \mathcal{C}'
  \;=\;
  \int_{T} e^{\mathcal{P}_2} \wedge \pi^\ast\left( e^{B_2} \wedge \mathcal{C}\right)\;,
\)
where $\pi$ is the projection from the fiber product of the given torus bundle with its T-dual
torus bundle and $\mathcal{P}_2$ is a canonical differential 2-form that this carries, called
the \emph{Poincar{\'e} 2-form}. Formally, T-duality  amount to this operation
exhibiting an isomorphism in H-flux twisted de Rham cohomology:
\(
  \xymatrix{
    H^{\bullet + \tilde H_3}( X^{10}_{\mathrm{IIA}} )
     \ar[rr]^{ \int_T e^{\mathcal{P}_2} \wedge \pi^\ast(-) }_{\simeq}
    &&
    H^{\bullet + \tilde H'_3}(  X^{10}_{\mathrm{IIB}} )
  }
\;.
\)
By higher generalization, there is a similar isomorphism on the $G_7$-flux
twisted real cohomology of the 3-sphere bundle over extended M-super-spacetime:
\(
 \label{eC6Twisted}
  e^{C_6} \wedge \mathcal{F}'
  \;=\;
  \int_{S^3} e^{\mathcal{P}_6} \wedge \pi^\ast\left( e^{C_6} \wedge \mathcal{F}\right)
  \,,
\)
where now $\pi$, $\pi'$ denote the two projections from the fiber product of the 3-sphere bundle with itself.
This yields an isomorphisms in twisted cohomology
\(
  \label{IsoIn7Twisted}
  \xymatrix{
    H^{\bullet + \tilde G_7}( Y^{11}_M )
     \ar[rr]_{\simeq}^{\;\; \int_{S^3} e^{\mathcal{P}_6} \wedge \pi^\ast(-) \;\;}
     &&
    H^{\bullet + \tilde G_7}(  Y^{11}_M )
  }
\;.
\)
as may be seen from the following argument.
The higher Poincar{\'e} form is $\mathcal{P}_6 = {\pi'}^\ast(C_3) \wedge \pi^\ast(C_3)$
and let us uniquely decompose any cochain as
\(
  \label{UniqueWrappingDecomposition}
  \mathcal{F} = \mathcal{F}_{\mathrm{nw}} + C_3 \wedge \mathcal{F}_{\mathrm{w}}\;,
\)
i.e., into the coefficients of its components that wrap ($\mathrm{w}$) or do not wrap ($\mathrm{nw}$) the 3-sphere fiber.
Consequently, we find that
{\small
\begin{equation}
  \label{PullTensorPushActionOnGeneralElement}
  \begin{aligned}
    &
    \int_{S^3} e^{ \mathcal{P}_6 }  \pi^\ast
    (\mathcal{F})
    \\
    & = \hspace{-2mm}
    \int_{S^3}
    \hspace{-2mm}
    \big(
      (
        1 +  {\pi'}^\ast(C_3) \wedge \pi^\ast(C_3)
      )
      \wedge
      (
        \pi^\ast(\mathcal{F}_{\mathrm{nw}})
        +
        \pi^\ast(C_3) \wedge \pi^\ast(\mathcal{F}_{\mathrm{w}})
      )
    \big)
    \\
    & =\hspace{-2mm}
    \int_{S^3}
      \pi^\ast(C_3)
      \wedge
      \big(
        \pi^\ast(\mathcal{F}_{\mathrm{w}})
        -
        {\pi'}^\ast( C_3 ) \wedge \pi^\ast( \mathcal{F}_{\mathrm{nw}} )
      \big)
      +
    \underset{ = 0}{
    \underbrace{
    \int_{S^3} \hspace{-2mm}
      \big(
        \cdots
      \big)
    }}
    \\
    & =
    - \mathcal{F}_{\mathrm{w}} +   C_3 \wedge \mathcal{F}_{\mathrm{nw}}
    \,.
  \end{aligned}
\end{equation}
}
This shows that the formula \eqref{eC6Twisted} acts as a linear isomorphism on cochains, given by exchanging
3-spherical wrapping modes with non-wrapping modes -- just as befits a higher T-duality.
To see that this indeed induces an isomorphism in 7-twisted
cohomology it is thus sufficient to check that this operation intertwines the 7-twisted differentials.
Straightforward computation reveals that this is the case precisely due to the special form \eqref{C3-T} of the 7-form flux!
For more mathematical details we refer the reader to \cite[Theorem 3.17]{Higher}.

Here we instead amplify the resulting physics:
The 7-twisted M-theoretic flux $e^{C_6} \wedge \mathcal{F}$,
which is thereby a new effect seen in M-theory, ought to couple to whatever it is on which
 M5-branes may end. This is a very interesting subject of current investigation.
Nevertheless, we can identify its string theory limit: We invoke Ho{\v r}ava-Witten theory
\cite{HoravaWitten96b} in order to restrict them to a 10-dimensional heterotic supergravity boundary of 11-dimensional superspacetime. On this boundary the M5-brane flux $\tilde G_7$
restricts to the NS5-brane flux $H_7$ and hence we deduce that the M-theoretic $\tilde G_7$-twisted flux
$e^{C_6} \wedge \mathcal{F}$ becomes in heterotic string theory an $H_7$-twisted flux
$e^{B_6} \wedge \mathcal{F}^{\mathrm{het}}$.

\vspace{-5mm}
\(
  \left.
  \begin{array}{c}
    \tilde G_7
    \\
    e^{C_6} \wedge \mathcal{F}
  \end{array}
  \right\rbrace
  \xymatrix{
    \ar@{|->}[rr]^{ {\begin{array}{c} \mbox{\tiny M/HET duality} \end{array}} }
    &&
  }
  \left\lbrace
  \begin{array}{c}
    H_7
    \\
    e^{B_6} \wedge \mathcal{F}^{\mathrm{het}}
  \end{array}
  \right.
\)
Precisely this had been identified in \cite{Sati09}: $\mathcal{F}^{\mathrm{het}}_2$ must
be the heterotic gauge field strength and $\mathcal{F}^{\mathrm{het}}_8$ its 10-dimensional
Hodge dual. The fact that these jointly form an $H_7$-twisted cocycle follows 
\cite{Sati09} from the Green-Schwarz anomaly cancellation
mechanism, i.e. from the twisted Bianchi identity  
$$
  d H \propto \mathrm{tr}\left( \mathcal{F}_2^{\mathrm{het}} \wedge \mathcal{F}_2^{\mathrm{het}} \right)
$$
for the NS 3-form flux.
This is, of course, the origin of all string theory unification.

{\scriptsize
\begin{center}
\begin{tabular}{|c||c|}
  \hline
  {\bf T-duality}
  &
  \begin{tabular}{c}
    {\bf Higher}
    {\bf T-duality}
  \end{tabular}
  \\
  \hline \hline
  \begin{tabular}{c}
    Torus bundle
    \\
    MC 1-form $\theta$
  \end{tabular}
  &
  \begin{tabular}{c|c}
  \begin{tabular}{c}
    3-sphere bundle
    \\
    MC 3-form $C_3$
  \end{tabular}
  &
  \begin{tabular}{c}
    517-torus bundle
    \\
    decomposable $C^{\mathrm{exc}}_3$
  \end{tabular}
  \end{tabular}
  \\
  \hline
  \begin{tabular}{c}
    String
    \\
    $H_3 + \theta \wedge F'_2$
  \end{tabular}
  &
  \begin{tabular}{c}
    M5-brane
    \\
    $G_7 + \tfrac{1}{2}C^{(\mathrm{exc})}_3 \wedge G_4$
  \end{tabular}
  \\
  \hline
  \begin{tabular}{c}
    D-branes
    \\
    $e^{B_2} \wedge \mathcal{C}$
  \end{tabular}
  &
  \begin{tabular}{c}
    Exotic M(9)-branes?
    \\
    $e^{C_6} \wedge \mathcal{F}$
  \end{tabular}
  \\
  \hline
\end{tabular}
\end{center}
}

One may ask more generally whether higher T-duals exist and are unique in our supersymmetric context. For bosonic (rational) spherical T-duality
this is discussed in \cite{LSW, BEM2}, but we amplify that here, using \cite{Higher}, we are considering
higher T-duality constructed via supersymmetry. We believe this question deserves to be fully addressed elsewhere.

\section{\bf  Exceptional generalized super-geometry}

Now we elucidate the subtlety that spherical T-duality of M5-branes acts not on internal spaces within
super-spacetime, but rather on the M2-gerbe \emph{over} the total spacetime. To do so, we ask for parameterization
of that super 2-gerbe by an \emph{ordinary} super-space, such that
it still serves as a target space for the M5-brane sigma-model.
$$
\hspace{-.4cm}
  \xymatrix@R=4pt@C=17pt{
   {
   \underbrace{\fbox
    {
    $
\arraycolsep=.2pt\def\arraystretch{-2}
      \begin{array}{c}
        \mbox{\scriptsize  Exceptional-generalized }
        \\
        \mbox{\scriptsize super-geometry}
      \end{array}
    $
    }}
    }
    &
       {
   \underbrace{ \fbox
    {
    $
    \arraycolsep=.2pt\def\arraystretch{-2}
    \begin{array}{cc}
      \mbox{\scriptsize  Super 2-gerbe}
      \\
     \mbox{\scriptsize  over super-spacetime}
    \end{array}
    $
    }}
    }
    \\
    \mathbb{R}^{10,1\vert\mathbf{32}}_{\mathrm{exc},s}
    \ar[r]^{ \phi_s }
    &
    \mathfrak{m}2\mathfrak{brane}
    \\
    \\
    \\
    \\
    \Sigma_{M5}
    \ar@{..>}@/_2pc/[uuuur]|{ \mbox{\it \tiny Sigma-model-}\atop \mbox{\it \tiny\& gauge-field } }
    \ar@{..>}@/^1pc/[uuuu]|{ \mbox{\it \tiny U-duality equivariant} \atop
    \mbox{\it \tiny sigma-model- \& gauge-field }  }
  }
$$
By the analysis in \cite{Higher} we may achieve this by re-interpreting the
``hidden'' DF-algebra of the supergravity literature \cite{DF, BAIPV04, ADR16}
as being the algebra of super-symmetric differential forms on this parameter
superspace, which thereby is identified as a super-geometric refinement
$\mathbb{R}^{10,1\vert \mathbf{32}}_{\mathrm{exc},s}$ of the
M-theoretic exceptional generalized geometry proposed in  \cite{Hull07b}:

\vspace{.3cm}

\hspace{-.4cm}
\scalebox{.9}{
$
    \mathbb{R}^{10,1\vert \mathbf{32}}_{\mathrm{exc},s}
    \simeq
    \underset{
      \mbox{\tiny Exceptional-generalized Super-spacetime}
    }{
    \underbrace{
    \underset{
      \mbox{\tiny Exceptional-generalized spacetime}
    }{
    \underbrace{
    \underset{
      \mbox{\tiny Super-spacetime}
    }{
    \underbrace{
    \underset{
      \mbox{\tiny Spacetime}
    }{
    \underbrace{
      \mathbb{R}^{10,1\vert \mathbf{32}}
    }}
    \oplus
    \mathbf{32}
    }
    }
    \oplus
    \Exterior^2(\mathbb{R}^{10,1})^\ast
    \oplus
    \Exterior^5(\mathbb{R}^{10,1})^\ast
    }
    }
    \oplus
    \mathbf{32}
    }
    }
$
}

\medskip
\noindent Thus we discover that the relevant parameterization renders the M5-brane
a plain (non-gauged) sigma-model on
exceptional spacetime, as recently proposed in \cite{Sakatani16}.
Under this identification, the comparison map $\phi_s$ pulls back the universal C-field $c_3$
to a decomposable form:
\(
  \label{Decompo}
  c_{\mathrm{exc},s} = \phi_s^\ast(c_3)
  \,.
\)
We observe that this decomposition is indeed what makes the idea of exceptional-generalized
geometry work: Each choice of section (linear splitting) of the exceptional tangent bundle
$$
  \raisebox{20pt}{
  \xymatrix{
    \mbox{\tiny Moduli space}
    \ar@{}[d]|{ \mbox{ \tiny Classifying map} }
    &
    \mathbb{R}^{10,1\vert \mathbf{32}}_{\mathrm{exc},s}
    \ar[d]^{\pi_{\mathrm{exc},s}}
    \\
    \mbox{\tiny Spacetime}
    &
    \mathbb{R}^{10,1\vert \mathbf{32}}
    \ar@/^1pc/[u]^{\sigma}
  }
  }
$$
allows to pull-back the universal
decomposed C-field and thus induce an actual C-field configuration satisfying the M2-brane
 super-torsion constraint
(\cite{BST2}):
\(
  \underset{
    \mbox{\tiny Torsion constraint}
  }{
  \underbrace{
  d \hspace{.1mm}
  \underset{
    { \mbox{\tiny C-field} \atop \mbox{\tiny configuration} }
   }{\underbrace{(\sigma^\ast c_{\mathrm{exc},s})}}  \hspace{-.5mm}=
   \tfrac{i}{2}\overline{\psi}\Gamma_{a b} \psi \wedge e^a \wedge e^b
  }}
  \,.
\)

\vspace{1mm}
This leads to the following characterizations:

{\it
\noindent {\bf 1.} Each of the fermionic extensions $\mathbb{R}^{10,1 \vert \mathbf{32}}_{\mathrm{exc},s}$
    of the
    exceptional super-spacetime $\mathbb{R}^{10,1\vert \mathbf{32}}$,
    for each nonzero real parameter $s$, serves as a moduli space for C-field configurations.

\noindent {\bf 2.} The decomposed $C$-field $c_{\mathrm{exc},s}$ in $\mathrm{CE}( \mathbb{R}^{10,1\vert \mathbf{32}}_{\mathrm{exc},s})$
   is the corresponding universal field on the moduli space, whose pullback along classifying maps $\sigma$ yield the
   actual C-field configurations on super-Minkowski spacetime.
}

\section{Transmutation of 3-spheres to 517-tori}

We next ask whether the spherical T-duality of M5-branes passes along the decomposition map
to the exceptional-generalized superspacetime. If we think of the latter as compactified, this means to
trade the original rational 3-sphere for a 517-torus ($517=528-11$)
with tangent space $\Exterior^2(\mathbb{R}^{10,1})^\ast \oplus \Exterior^5(\mathbb{R}^{10,1})$; see the picture on p. 2.

Intuitively it might seem that a duality embodied by
exchanging 3-spherical wrapping/non-wrapping modes has no chance to be retained if the 3-sphere is replaced
by a (high-dimensional) torus. Concretely, the issue is that the
unique wrapping/non-wrapping decomposition of \eqref{UniqueWrappingDecomposition} is used in the proof
of the duality \eqref{PullTensorPushActionOnGeneralElement}, which does generally hold for actual 3-sphere bundles
but not necessarily for higher torus bundles. But with \cite{Higher} we may observe that this uniqueness of 5-brane wrapping modes is mathematically expressed by the
cohomology of the operation of taking the wedge product with the decomposed C-field (\ref{Decompo}). In \cite{Higher} this is called the
\emph{C-cohomology} of the exceptional generalized super-spacetime:
\(
  \label{CCohomology}
  \mbox{C-cohomology}
  =
  \frac{
    \mathrm{kernel}( c_{\mathrm{exc},s} \wedge(-) )
  }{
    \mathrm{image}( c_{\mathrm{exc},s} \wedge(-) )
  }
  \,.
\)
An argument in homological algebra computations reveals that the criterion for
3-spherical T-duality to be retained on exceptional super-spacetime is that the $\mathrm{Spin}$-invariant C-cohomology of the
decomposed C-field \emph{vanishes}.

Concrete computation shows \cite{Higher} that the C-cohomology is concentrated on invariant fermionic multiples of
the volume form $\mathrm{vol}_{528}$ of the
528-dimensional exceptional tangent bundle. We indicate how this comes about:
First, we may write the decomposed C-field as the sum of a bosonic and a bifermionic contribution:
\(
  c_{\mathrm{exc},s} = (c_{\mathrm{exc},s})_{\mathrm{bos}} + (c_{\mathrm{exc},s})_{\mathrm{ferm}}
  \,.
\)
Now from \cite{DF, BAIPV04, ADR16} one finds that the fermionic part turns out to be an \emph{odd-symplectic form} of the schematic form
\(
  (c_{\mathrm{exc},s})_{\mathrm{ferm}}
  =
  (\overline{\eta}_\alpha \wedge \psi_\beta)  \Gamma_A^{\alpha \beta} E^A
\)
for $(E^A) = (e^a , B_{a_1 a_2}, B_{a_1,\cdots, a_5})$ the exceptional-generalized vielbein.
Then an argument that builds on \cite{Sev05} shows that already the  C-cohomology of the bifermionic component is concentrated on the
invariant combinations of $\psi$ times the 528-volume form.
If we introduce on supersymmetric forms the bigrading
by bosonic form-degree $n_{\mathrm{bos}}$ and fermionic form-degree $n_{\mathrm{ferm}}$ then
these classes correspond to the dots in the following diagram:

\vspace{-1.5cm}
\scalebox{.7}{
  $$
    \xymatrix{
      & & & &&&
      \\
      & & & &  &&
      \\
      & & & &&  & &
      \\
      & & &  \ddots &
      \\
      &
      &
      &
      & {}_{\mathllap{(522,6)}}\bullet
       \ar[r]|{d_1}
       \ar[rrd]|{d_2}
       \ar[rrrdd]|{d_3}
       &
      \\
      &
      & &&& {}_{\mathllap{(525,3)}}\bullet &&&&&&
      \\
      \ar[rrrrrrrr]_>{ n_{\mathrm{\bf bos}} - n_{\mathrm{\bf ferm}}/2 }_>>>>>>>>>>>>>>>>>>>>>>>>>>{(528,0)} && &&&& \bullet  &&&  &&
      \\
      &  \ar[uuuuu]^>{ n_{\mathrm{\bf ferm}}/2 } & &&& &&&
    }
  $$
}
Finally, a standard argument from homological algebra (a ``double complex spectral sequence'') shows that the C-cohomology of the full
decomposed C-field is obtained from that of its bifermionic summand by
passing to the joint cohomology of an infinite sequence of higher differentials $d_r$, as indicated in the diagram. Since
in the present case none of these
differentials can go between two non-trivial classes (two dots in the diagram) there is in fact no further correction
(``the spectral sequence collapses'') and hence the claimed result follows.

This shows that 3-spherical T-duality of the M5-brane does indeed
pass to the 517-toroidal exceptional-generalized supergeometry,
except possibly for a slight ``fuzziness'' if $\tilde G_7$-twisted cocycles contain a summand proportional to
$\mathrm{vol}_{528}$. It seems plausible that this is not the case for any non-trivial $\tilde G_7$-twisted cocycle,
but we leave this as an open mathematical question.

\section{\bf Toroidal T-duality on exceptional super-spacetime}

With a new M-theoretic kind of higher T-duality on exceptional-generalized super-spacetime
$\mathbb{R}^{10,1\vert \mathbf{32}}_{\mathrm{exc},s}$ thus established,
it is curious to observe that, in fact, there is also string-theoretic 517-toroidal T-duality on
the exceptional coordinates of $\mathbb{R}^{10,1\vert \mathbf{32}}_{\mathrm{exc},s}$
\emph{over} 11d super-spacetime.

By \cite{FSS16}, this is induced by the presence of a a super-3-cocycle of the form of a
T-dualizable H-flux (\ref{H3-T}), and inspection \cite{Higher} shows that such may be extracted from the
main spinorial Fierz-identity \cite{DF, BAIPV04, ADR16} that controls the existence of the exceptional-generalized supergeometry. It is of the following form:
\(
   \label{StringCocycleOnExceptionalSpacetime}
    \begin{aligned}
  \mu_3 =   (s+1)
     e_a \wedge \overline{\psi} \Gamma^a \psi
      &-
     B_{a_1 a_2} \wedge \overline{\psi} \Gamma^{a_1 a_2} \psi &
     \\
     +
     (1 + \tfrac{s}{6}) &
     B_{a_1 \cdots a_5} \wedge \overline{\psi} \Gamma^{a_1 \cdots a_5} \psi
     \,.
    \end{aligned}
\)
This reveals that there is a large exotic  self-duality structure on the ``exceptional string sigma-model'' recently discussed
in \cite{ArvanitakisBlair18}.

\section{\bf  Parity symmetry in M-theory}

A famous exotic symmetry of 11-dimensional supergravity is ``parity symmetry'',
which says that the theory is invariant under an odd number of spacetime-reflections if these are accompanied
by sending the C-field to its negative \cite{DLP}
\(
  C_3 \mapsto - C_3
  \,.
\)
We may observe \cite{Higher} that this operation lifts to an equivalence $\rho^\ast$ of M5-brane 7-flux-twisted cohomology
on exceptional-generalized super-spacetime, on par with the spherical T-duality equivalence \eqref{IsoIn7Twisted}.
For this we declare that under spacetime-reflection the exceptional-generalized super-vielbein transforms in the
following curious way:

The ordinary vielbein $e^a$ and the M5-wrapping modes $B^{a_1 \cdots a_5}$ pick up a sign precisely
when their indices \emph{do}
contain a given direction being reflected, while the M2-brane wrapping modes $B^{a_1 a_2}$ pick up a sign precisely if their indices
\emph{do not}.

Concrete computation shows that this yields a  self-equivalence of the same form as (\ref{IsoIn7Twisted}):
\(
     H^{\bullet + \tilde \mu_{{}_{M5,s}}}( \mathbb{R}^{10,1\vert\mathbf{32}}_{\mathrm{exc},s} )
     \xrightarrow[\simeq]{\rho^\ast}
    H^{\bullet + \tilde \mu_{{}_{M5,s}}}( \mathbb{R}^{10,1\vert\mathbf{32}}_{\mathrm{exc},s} )
\;.
\)
This means that we should view parity and 3-spherical T-duality as generators that \emph{jointly}
 induce a larger M-theoretic duality group which also contains their composite operation:
\(
  \xymatrix{
    &
    H^{\bullet + \tilde \mu_{{}_{M5}}}
    \ar[dr]_{\simeq}^{ \rho^\ast }
    \\
    H^{\bullet + \tilde \mu_{{}_{M5}}}
    \ar[ur]_{\simeq}^{\hspace{-7mm} \int_{S^3} e^{\mathcal{P}_6} \wedge \pi^\ast(-)  }
    \ar[rr]^\simeq_{
      \mbox{\tiny `Paritized'}
      \atop
      \mbox{\tiny 3-spherical T-duality }
    }
    &&
    H^{\bullet + \tilde \mu_{{}_{M5}}}\;.
  }
\)
Indeed, this is compatible with the results of parity in the topological sector in \cite{DFM}.
%
%
%

Theories describing multiple membranes
should preserve parity, which places constraints on the structure
constants appearing in the Lagrangian \cite{BLMP}.
Chern-Simons-matter theories describing fractional M2-branes \cite{ABJ},
 arising from backgrounds
${\rm AdS}_4 \times S^7/\Z_k$ with torsion class
in $H^4(S^7/\Z_k; \Z)\cong \Z_k$, lead to symmetries
of the form $U(n+ \ell)_k \times U(n)_{-k}$, where $k$ is the
Chern-Simons level. The parity symmetry acting as
$C_3 \mapsto -C_3$ takes $k$ to $-k$ and then
$G_4 \mapsto k - G_4$, so that the rank $\ell$ goes to
$\ell-k$ in structure groups of level $-k$ \cite{ABJ}.
Our results seem to place these theories in the context of our higher T-duality, but
we leave this for future investigation.

From global geometric and topological perspective,
 M-theory is parity invariant, and so should in principle be formulated
 in a way that makes sense on unoriented, and possibly
 orientable manifolds (see \cite{DFM}). We hope that our formulation
 provides some insight into this problem.

Now we consider the degree four field to be captured by an $E_8$ bundle
over spacetime, as in \cite{Wi1}\cite{DMW} \cite{DFM}, characterized by
an integral degree four class $a$.
When the Pontrjagin class is zero, the parity transformation acts
on the degree four classes as
$
a \mapsto - a
$ and
$G_4 \mapsto -G_4
$.
Note that $E_8$ and ${\rm SU}(2)$ have
the same topology (namely, rational homotopy and cohomology)
in the range of dimensions of M-theory, so that
 we view our $E_8$
bundle over an 11-dimensional base space as a 3-sphere bundle.
Taking two such bundles $E$ and $E'$ with  classes  $a$ and $-a$,
respectively,  then puts the two bundles as a parity dual pair, which
fits into our discussion of T-duality for rational sphere bundle as a special
case.
A parity-invariant formulation of the $E_8$ model is given in \cite{DFM}
 by passing to the orientation double
cover $Y_d$ of spacetime $Y$
which  we could use here.

%
%
%
%
%

%



\vspace{-.3cm}

\begin{thebibliography}{99}

\bibitem{Vafa05}
C. Vafa, Einstein Symposium 2005,
[hep-th/0509212].

\bibitem{Polchinski94}
J. Polchinski,
 Phys. Rev. D {\bf 50}, 6041 (1994), [hep-th/9407031].

 \bibitem{BeckerBeckerStrominger95}
 K. Becker, M. Becker, A. Strominger,
 Nucl. Phys. {\bf B 456}, 130 (1995),
[hep-th/9507158].

\bibitem{Bu1}
T. H. Buscher,
 Phys. Lett. {\bf B194}, 59 (1987) .

\bibitem{Bu2}
T. H. Buscher,
Phys. Lett. {\bf B201}, 466 (1988).

\bibitem{GPR}
A. Giveon, M. Porrati, E. Rabinovici,
Phys. Rept. {\bf 244},  77 (1994),
hep-th/9401139.

\bibitem{Duff99}
M. Duff,
{\it The world in 11 dimensions: Supergravity, Supermembranes and M-theory},
IoP 1999

\bibitem{GV1}	
R. Gopakumar, C. Vafa
Adv. Theor. Math. Phys. {\bf 3}, 1415 (1999),
[hep-th/9811131].

\bibitem{GV2}	
R. Gopakumar, C. Vafa,
[hep-th/9809187].



\bibitem{AETW87}
A. Ach{\'u}carro, J. Evans, P. Townsend, D. Wiltshire,
Phys. Lett. {\bf B 198}, 441 (1987).

\bibitem{AzTo89}
J. de Azc{\'a}rraga, P. Townsend,
Phy. Rev. Lett. {\bf 62},  2579 (1989).



\bibitem{Higher}
D. Fiorenza, H. Sati, U. Schreiber,
[1803.05634].

\bibitem{FSS16}
D. Fiorenza, H. Sati, U. Schreiber,
[1611.06536].

\bibitem{BEM}
P. Bouwknegt, J. Evslin, V. Mathai,
Phys. Rev. Lett. {\bf 92}, 181601 (2004),
[hep-th/0312052].



\bibitem{CL}
A. Candiello, K. Lechner,
	Nucl. Phys. {\bf B412}, 479  (1994),
[hep-th/9309143].


\bibitem{Howe97}
P. Howe,
Phys. Lett. {\bf B 415}, 149 (1997),
[hep-th/9707184].

\bibitem{BST1}
E. Bergshoeff, E. Sezgin, P. K. Townsend,
Phys. Lett. {\bf 169B}, 191 (1986).

\bibitem{BST2}
E. Bergshoeff, E. Sezgin, P. K. Townsend,	
 Phys. Lett. {\bf B189}, 75  (1987).

\bibitem{BLNPST97}
I. Bandos, K. Lechner, A. Nurmagambetov, P. Pasti, D. Sorokin, M. Tonin,
Phys. Rev. Lett. {\bf 78},  4332 (1997),
[hep-th/9701149].

\bibitem{LSW}
J. A. Lind, H. Sati, C. Westerland,
[1601.06285].


\bibitem{BEM2}
P. Bouwknegt, J.  Evslin, V. Mathai,
Commun. Math. Phys. {\bf 337}, 909 (2015),
[1405.5844].

\bibitem{Hull07a}
C. M. Hull,
JHEP0710:057, 2007
[hep-th/0604178]


\bibitem{FSS13}
D. Fiorenza, H. Sati, U. Schreiber,
IJGMMP {\bf 12},  1550018 (2015),
[1308.5264].


\bibitem{Hull07b}
C. Hull,
JHEP {\bf 0707}, 079  (2007),
[hep-th/0701203].

\bibitem{FSS15}
D. Fiorenza, H. Sati, U. Schreiber,
J. Math. Phys. {\bf 56}, 102301 (2015),
[1506.07557].


\bibitem{DF}
R.~D'Auria, P.~Fr{\'e},
\newblock Nucl. Phys. {\bf B} 201, 101 (1982),

\bibitem{Hori}
K. Hori,
Adv. Theor. Math. Phys. {\bf 3}, 281 (1999),
[hep-th/9902102].

\bibitem{T}
D. Fiorenza, H. Sati, U. Schreiber,
[1712.00758].


\bibitem{HoravaWitten96b}
P. Ho{\v r}ava, E. Witten,
Nucl. Phys. {\bf B475}, 94 (1996),
[hep-th/9603142].

\bibitem{Sati09}
H. Sati,
J. Geom. Phys. {\bf 59}, 369 (2009),
[hep-th/0701232].


\bibitem{ADR16}
L. Andrianopoli, R. D'Auria, L. Ravera,
JHEP {\bf 1608}, 095  (2016),
[1606.07328].


\bibitem{BAIPV04}
I.A. Bandos, J.A. de Azcarraga, J.M. Izquierdo, M. Picon, O. Varela,
Phys. Lett. {\bf B596}, 145 (2004),
[hep-th/0406020].


\bibitem{Sakatani16}
Y. Sakatani, S. Uehara,
Phys. Rev. Lett. {\bf 117}, 191601  (2016), [1607.04265].


\bibitem{Sev05}
P. {\v S}evera,
Lett. Math. Phys {\bf 78}, 55  (2006),
[math/0506331].

\bibitem{ArvanitakisBlair18}
A. Arvanitakis, C. Blair,
[1802.00442].


\bibitem{DLP}	
M. J. Duff, B. E. W. Nilsson, C. N. Pope,
Phys. Rept. {\bf 130}, 1 (1986).

\bibitem{ABJ}
O. Aharony,
 O. Bergman, D. L. Jafferis,
 JHEP  {\bf 0811}, 043  (2008),
[0807.4924].

\bibitem{BLMP}
J. Bagger, N. Lambert, S. Mukhi, C. Papageorgakis,
Phys. Rept. {\bf 527}, 1  (2013),
[1203.3546].



\bibitem{Wi1}
E. Witten,
	J. Geom. Phys. {\bf 22}, 103 (1997),
[hep-th/9610234].

\bibitem{DMW}
D.-E. Diaconescu, G. Moore, E. Witten,
Adv. Theor. Math. Phys. {\bf 6},  1031 (2003),
[hep-th/0005090].

\bibitem{DFM}
D.-E. Diaconescu, D. Freed,  G. Moore,
London Math. Soc. Lecture Note Ser., vol. 342,
 Cambridge, 2007, pp. 44-88,
[hep-th/0312069].
















%



\end{thebibliography}
\end{document}